\documentclass[
reprint,
superscriptaddress,
amsmath,amssymb,
aps,
]{revtex4-1}

\usepackage{color}
\usepackage[dvipdfmx]{graphicx}% Include figure files
\usepackage[FIGTOPCAP]{subfigure}
\usepackage{dcolumn}% Align table columns on decimal point
\usepackage{dcolumn}% Align table columns on decimal point
\usepackage{bm,braket}% bold math
\usepackage{comment}
\usepackage{ulem}%取り消し線使うため
\usepackage{here}
\usepackage{subcaption} % サブキャプション対応

\usepackage{titlesec}

\titleformat{\subsubsection}[block]  % [block]は改行を意味する
  {\normalfont\bfseries}            % フォントスタイル（通常フォント + 太字）
  {\textit{\thesubsubsection}}      % 番号部分を斜体に
  {1em}                             % 番号とタイトルの間隔
  {}                                % タイトル後の装飾なし

\usepackage{hyperref} % ハイパーリンク用パッケージ

\begin {document}
%section {title}
%\preprint{APS/123-QED}

\title{Unveiling Optimal Diffusion for Infection Control in Brownian Particle Systems}% Force line breaks with \\

\author{Kaito Takahashi}
\affiliation{%
	Department of Physics and Astronomy, Tokyo University of Science, Noda, Chiba 278-8510, Japan
}%

\author{Makiko Sasada}
\email{sasada@ms.u-tokyo.ac.jp}
\affiliation{%
	Graduate School of Mathematical Sciences, University of Tokyo, 3-8-1, Komaba, Meguro-ku, Tokyo, 153-8914, Japan
}%

\author{Takuma Akimoto}
\email{takuma@rs.tus.ac.jp}
\affiliation{%
	Department of Physics and Astronomy, Tokyo University of Science, Noda, Chiba 278-8510, Japan
}%

%}


\date{\today}% It is always \today, today,

\begin{abstract}
Understanding the spread of infectious diseases requires integrating movement, physical constraints, and spatial configurations into epidemiological models. In this study, we investigate how particle diffusivity, hardcore interactions, and non-equilibrium initial conditions influence infection dynamics within a system of Brownian particles. {\color{black}Using numerical simulations and theoretical analysis, we reveal a nontrivial relationship between diffusivity and the speed of infection spread. Specifically, when particles are initially positioned at uniform distances greater than the infection radius---a non-equilibrium configuration---there exists an optimal diffusion coefficient that minimizes the infection propagation speed.} This counterintuitive result arises from  the competition between diffusive timescales and the rate of infection transmission.  The presence of an optimal diffusivity is observed both in systems with and without hardcore interactions, provided that the infection radius exceeds the mean lattice spacing.  Our findings provide a theoretical framework for understanding and controlling the spread of infections in confined and diffusive environments, with potential implications for designing movement-based strategies for infection control.
\end{abstract}

\maketitle

\section{Introduction}

Mathematical models for understanding and predicting the spread of infectious diseases play a fundamental role in epidemiology \cite{choisy2007mathematical}. One of the most classical and widely-used models is the Susceptible-Infected-Recovered (SIR) model, proposed by Kermack and McKendrick in 1927 \cite{kermack1927contribution}. This model divides the population into three compartments—susceptible, infected, and recovered—and describes the dynamics of disease spread based on interaction rates among these groups. The SIR model, along with its variants (SIS, SEIR, etc.), has been applied to study various infectious diseases, including influenza and COVID-19 \cite{anderson1991infectious,hernandez2014effects,he2020seir,bartlett2016mathematical,side2013sir}. However, these compartmental models often overlook spatial interactions and individual mobility patterns, which are critical factors in realistic simulations of infectious disease dynamics \cite{cooper2020sir, allen1994some}.

To address these limitations, spatial models have been developed to incorporate local interactions and individual movement patterns \cite{hethcote2000mathematics, funk2010modelling, shibuya2010infection}. For instance, the spatial SIR model assumes that infections occur only within a defined spatial range, creating heterogeneous infection patterns that depend on spatial constraints. Recently, particle-based models have attracted attention for their ability to simulate infection dynamics in mobile populations, where individual movement influences transmission \cite{rodriguez2019particle,rodriguez2022epidemic,laskowski2011agent,silva2020covid,Castro2023}. These models are based on physical principles of particle dynamics and allow for a detailed analysis of how contact patterns and movement affect disease spread.

Recent developments in active matter physics provide a framework for understanding collective motion in non-equilibrium systems. Active matter systems, composed of self-propelling particles, exhibit emergent behaviors such as flocking and clustering due to local interactions \cite{Bechinger2016}. Notably, the Vicsek model demonstrates how alignment interactions lead to coherent group behavior \cite{Vicsek1995}, and this framework has been adapted to simulate infection dynamics %by replacing particle spin with infection states 
\cite{norambuena2020understanding, centres2024diffusion, cai2019stochastic}.

Certain active matter systems exhibit long-range displacements that can be modeled using L\'evy  walks, which are widely applied in ecological and epidemiological contexts to describe animal search strategies \cite{viswanathan1996levy, bartumeus2005animal} and pathogen spread \cite{keeling2011modeling, abhignan2021simulations, matthaus2011origin}. However, our study focuses on Brownian motion \cite{einstein1905motion,shibuya2010infection,schweitzer2003brownian}, which is isotropic and short-ranged, making it more suitable for investigating infection dynamics in confined environments. Additionally, we consider a one-dimensional system, which is relevant for real-world scenarios such as infection spreading in narrow corridors, queues, and transportation aisles, as well as pathogen transmission in microfluidic or biological channels \cite{derjany2020multiscale, li2020effects, figueroa2020coli}, where movement is effectively restricted to a linear geometry. The constraints imposed by diffusivity and hardcore interactions in such environments play a crucial role in shaping infection dynamics.
By examining the role of diffusivity and hardcore interactions in Brownian motion, our work bridges the gap between particle-level dynamics and macroscopic infection patterns, complementing studies in active matter systems.

%Certain active matter systems also incorporate L\'evy walk-like dynamics \cite{Zaburdaev2015}, characterized by intermittent long-range displacements. L\'evy walks, with their heavy-tailed displacement distributions, are widely used to model large-scale movement patterns in ecological and epidemiological contexts \cite{keeling2011modeling}, such as animal search strategies and pathogen spread \cite{viswanathan1996levy, bartumeus2005animal}. While L\'evy walks and active matter share similarities, especially in systems with intermittent or correlated motion, our study focuses on Brownian motion \cite{einstein1905motion}. Unlike L\'evy walks, Brownian motion is isotropic and short-ranged, making it better suited for investigating infection dynamics in confined environments.

In this study, we model infectious disease dynamics using  $N$  Brownian particles, where each particle can be in either an infected or susceptible state, mimicking spin dynamics on moving lattice points. Specifically, we  examine how the diffusion coefficient of the Brownian particle and infection radius affect the progression of infection. Additionally, we analyze how hardcore repulsive interactions between particles influence infection dynamics, taking into account the presence or absence of hardcore repulsive interactions. 
Our primary goal is to investigate the existence of an optimal diffusion coefficient that minimizes the infection spreading speed and to characterize how it arises from the interplay between diffusive timescales and the infection rate. Through numerical simulations and theoretical analysis, we aim to clarify how infection progression and spreading patterns are shaped by diffusive transport, spatial constraints, non-equilibrium initial configuration, and particle interactions.
{\color{black}We particularly focus on structured non-equilibrium initial conditions, where particles are uniformly spaced. While idealized, such configurations may arise in systems with spatial constraints or periodic organization. We also discuss scenarios where this structure is maintained or disrupted dynamically, highlighting its relevance to real-world settings.}

\section{Model}

We investigate infection propagation in a one-dimensional system containing $N$ Brownian particles under periodic boundary conditions with a system length \(L\). Initially, the particles are evenly spaced with an inter-particle distance of \(L/N\). In later sections, we explore how alternative initial conditions affect the infection dynamics. The motion of each particle follows standard Brownian dynamics, governed by the following stochastic differential equation \cite{uhlenbeck1930theory,nulton1993correlation}:
\begin{equation}
    \dot{x}_i(t) = \xi(t)
    \label{eq: BD}
\end{equation}
where $x_i(t)$ is a position of $i$-th Brownian particle, \(\xi(t)\) represents Gaussian white noise with zero mean, characterized by a diffusion coefficient \(D\) that governs the particle's mobility. The noise \(\xi(t)\) satisfies
\begin{equation}
\langle \xi(t) \rangle = 0, \quad \langle \xi(t)\xi(t') \rangle = 2D\delta(t - t'),
\end{equation}
where \(\langle \cdot \rangle\) denotes the ensemble average, and \(\delta(t - t')\) is the Dirac delta function.

\subsection{ Effect of Hardcore Repulsive Interaction}

We examine two scenarios: hardcore repulsive and free-diffusive Brownian particles. In the hardcore repulsive case, particles experience a hardcore repulsive interaction, which prevents them from passing through each other. {\color{black}In one dimension, this leads to a fixed spatial ordering: if a particle is labeled as $i$, the particle labeled $i+1$ is always located to its right.}  Due to the no-overtaking constraint, the sequence of particles remains unchanged over time, even though their individual positions fluctuate due to Brownian motion. This phenomenon, known as single-file diffusion in one-dimensional systems, ensures that particle order remains fixed during motion \cite{kollmann2003single, lutz2004single}. As a result, neighboring particles are unable to overtake each other, and their spatial order is maintained throughout the dynamics, preventing overlap \cite{percus1974anomalous}. Consequently, single-file diffusion leads to sub-diffusive behavior, where the mean square displacement (MSD) grows as \(t^{1/2}\) instead of the linear growth in normal diffusion \cite{harris1965diffusion,gupta1995evidence,hahn1996single,kukla1996nmr,karger1992straightforward}. This constraint plays a significant role in modulating infection propagation, as infection can only spread sequentially among neighboring particles.

\subsection{Infection Propagation Model and Infection Probability}

In this study, we adopt the Susceptible-Infected (SI) model to describe the process of infection propagation \cite{bodo2016sis}. The SI model is particularly suited for understanding the early-stage dynamics of disease spread, where recovery or reinfection effects are negligible. By focusing solely on infection transmission, we can isolate the fundamental role of diffusion and spatial constraints in shaping the spread of disease. This allows us to examine how infection events are influenced by particle movement rather than static contact networks, avoiding additional complexity from recovery mechanisms.

Each particle (agent) possesses one of two states: ``susceptible" or ``infected." A susceptible particle may become infected if it is located within the infection radius \(r\) of an infected particle. When particle interactions are present and a single-file spatial constraint is imposed, hardcore repulsion governs infection propagation. In this case, infection can only spread sequentially to adjacent particles due to the hardcore repulsion constraint, which prevents particles from passing each other and interacting beyond immediate neighbors. As a result, infection cannot leap over other particles, even if the susceptible particle lies within the infection radius  $r$.  On the other hand, in free-diffusive systems, infection can leap over adjacent particles. In this scenario, even if a susceptible particle is separated by multiple neighbors, it can still become infected if it lies within the infection radius \(r\). This distinction results in  different infection propagation patterns and speeds between single-file systems and free-diffusive systems.

As an initial condition for infection, a single infected particle is placed at the center of the agent distribution, while all other particles are susceptible. The state of infection for the \(j\)-th particle at time \(t\) is denoted by \(s_j(t)\), where \(s_j(t) = 1\) indicates the particle is infected, and \(s_j(t) = 0\) indicates it is susceptible. This formulation follows the standard SI model \cite{bodo2016sis,bianconi2017epidemic}. 

To determine whether particle  $i$  is within the infection radius of particle  $j$, we introduce the indicator function  $\sigma_{ij}(t)$, which is defined differently for free-diffusive and hardcore repulsive systems.
In the free-diffusive case, the infection can spread to any susceptible particle within the infection radius  $r $. The indicator function  $\sigma_{ij}(t)$  is defined as
\begin{equation}
\sigma_{ij}(t) = 
\begin{cases} 
    1 & \text{if } {\color{black}\min \left( |\Delta x_{ij} (t)|, L - |\Delta x_{ij} (t)| \right) < r,} \\ 
    \\
    0 & \text{otherwise},
\end{cases}
\end{equation}
{\color{black}where $\Delta x_{ij} (t) = x_i (t) - x_j(t)$. This formulation accounts for periodic boundary conditions by considering the minimum distance between particles either directly or across the system boundary.}  Since particles are free to move without constraints, infections can leap over other particles, leading to faster transmission dynamics compared to the hardcore repulsive case.
In hardcore repulsive systems, where particles experience hardcore repulsive interactions, the indicator function  $\sigma_{ij}(t)$  is modified to restrict infections to nearest neighbors:
\begin{equation}
\sigma_{ij}(t) = 
\begin{cases} 
    1 & \text{if } |\Delta x_{ij} (t)| < r~\text{and } |i-j|=1, \\ 
    0 & \text{otherwise}.
\end{cases}
\end{equation}
This modification implies that in hardcore repulsive systems, infections cannot leap over neighboring particles. The hardcore interaction restricts the spatial propagation of infection, leading to a slower spreading process compared to the free-diffusive case. Since single-file diffusion prevents overtaking, the infection is confined to nearest-neighbor transmission rather than spreading freely to any particle within radius  $r $.

The infection propagation follows a Poisson process \cite{daley1999epidemic,ball2002general}, where the probability of a susceptible particle $i$ becoming infected during the interval $[t, t+\Delta t]$ is given by
\begin{equation}
P_{infected}=1 - \exp \left( -\beta \sum_{j=1}^{N} \sigma_{ij}(t) s_j(t) \Delta t \right),
\end{equation}
assuming that  the value of $\sigma_{ij}(t)$ remains constant within  \([t, t+\Delta t]\). Here,  \(\beta\) represents the infection rate.  This expression depends on the number of infected neighboring particles and the susceptible particles within the infection radius.  To simplify this expression, in the limit of \(\Delta t \to 0\), where \(\beta \sum_{j=1}^{N} \sigma_{ij}(t) s_j(t) \Delta t \ll 1\), the infection probability can be approximated as
\begin{equation}
P_{infected} \approx \beta \sum_{j=1}^{N} \sigma_{ij}(t) s_j(t) \Delta t.
\end{equation}
This linear approximation highlights that the infection probability depends on the number of infected neighboring particles, the infection rate, and the time step \(\Delta t\). While particle motion is continuous, the simulation discretizes time into sufficiently small intervals to ensure accurate infection dynamics. Choosing an appropriately small  $\Delta t$  prevents numerical artifacts and ensures the validity of the approximation.
For hardcore repulsive systems, the hardcore interaction restricts $\sigma_{ij}(t)$ to nonzero values only for adjacent particles ($|i - j| = 1$), preventing the infection from leaping over intermediate particles. In contrast, free-diffusive systems allow the infection to spread freely, with $\sigma_{ij}(t)$ determined solely by the spatial distance $|x_i(t) - x_j(t)|$.

\subsection{Simulation Process}
{\color{black}In this study, we primarily initialize the system with a uniformly spaced (lattice) configuration. This structured, non-equilibrium setup enables us to investigate how initial spatial order influences infection dynamics. Although idealized, such configurations can emerge in confined or regularly organized environments. In later sections, we also examine how relaxing this spatial structure alters the infection spreading behavior.}
The simulation proceeds in discrete time steps of  $\Delta t $, iterating between two phases to simulate infection propagation and particle movement: 

\begin{enumerate}
\item \textbf{Infection phase}: infection dynamics are discrete in state, meaning that each particle's state is either infected or susceptible and is updated synchronously across the system. If a susceptible particle lies within the infection radius of an infected particle, it becomes infected with probability  \(P_{infected}=\beta \sum_{j=1}^{N} \sigma_{ij}(t)  s_j(t) \Delta t\). This probability is calculated independently for each particle, and the procedure is repeated for all particles in the system during each time step. 

    \item \textbf{Particle movement phase}: Particle movement takes place in a continuous spatial domain and follows Brownian dynamics, as described in Eq.~(\ref{eq: BD}). In systems with hardcore interactions, particles move while enforcing the hardcore repulsion constraint, which ensures that no overlaps occur and that the relative order of particles is preserved, consistent with single-file diffusion. 
    In numerical simulations, each particle first undergoes an initial Brownian displacement before enforcing interactions. If this displacement does not violate the hardcore repulsion constraint, it is accepted. However, if the move would cause an overlap, the particle positions are adjusted accordingly to maintain single-file ordering, preventing overtaking.

\end{enumerate}
By alternating between these two phases, we simulate the interaction between diffusion and infection, allowing us to explore how the diffusion coefficient and particle interactions influence the infection propagation speed.

\section{Results}

\begin{figure*}
    \includegraphics[width=1.\textwidth]{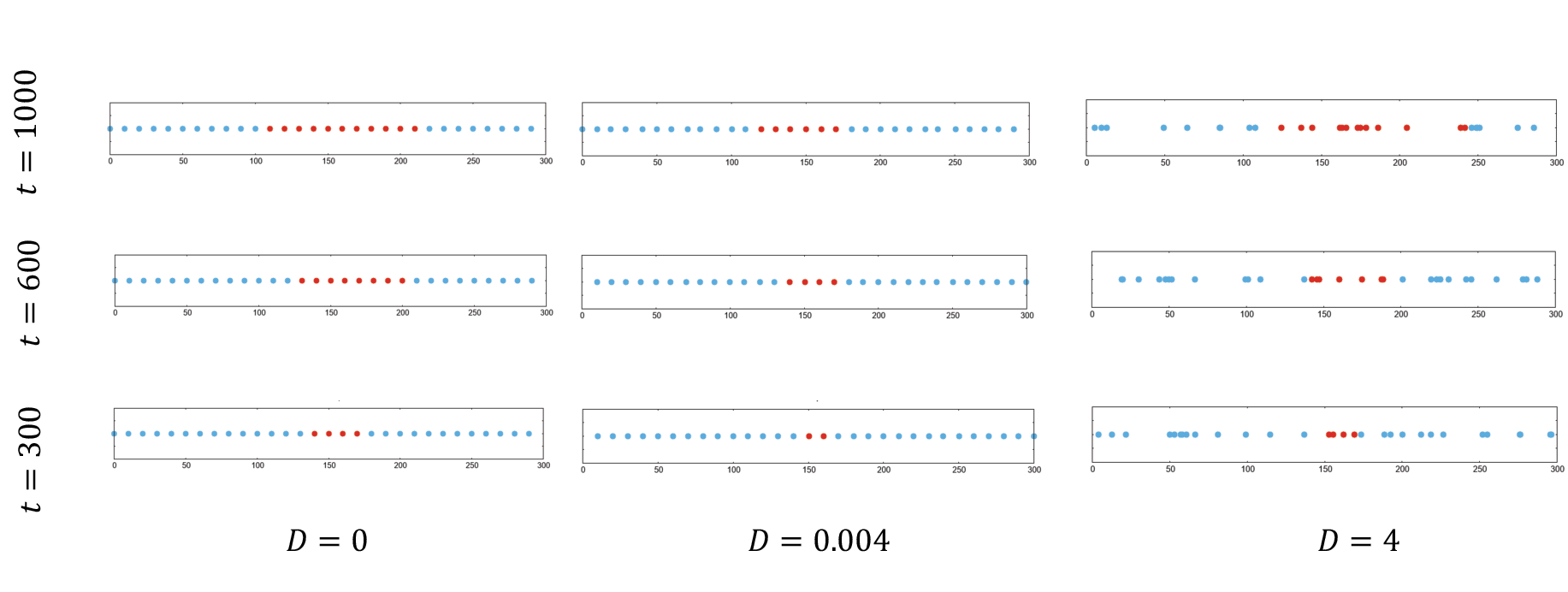} 
    \caption{ 
    Time evolution of the infected particles for diffusion coefficients ($D=0, 0.004$, and 4). The infection rate is \(\beta = 0.05\), the infection radius is $r=11$, the system size is $L=300$, and the number of particles is $N=30$. The initial particle configuration is a lattice arrangement.
       Red symbols represent infected particles, while blue symbols indicate susceptible ones. The infection speed is strongly dependent on the diffusivity. At $D = 0.004$, the spread of infection is most suppressed.
        }
    \label{fig:snapshots}
\end{figure*}

%\subsection{Evaluation of Relaxation Time and Infection Speed}
We investigate how the diffusivity of particles and the infection radius affect the infection speed in systems of $N$ Brownian particles with and without interactions, where the initial particle configuration is a lattice arrange. Figure~\ref{fig:snapshots} shows snapshots of infected and susceptible particles at different times. As time progresses, the infection spreads; however, the spatial patterns of infection vary significantly depending on the diffusivity, {\color{black}demonstrating the strong influence of mobility on infection propagation.}

{\color{black}We numerically observe that the average number of infected particles, $I(t)$, increases linearly with time at early times, although the precise shape of  $I(t)$  varies  depending on the diffusion coefficient  $D$.}
To evaluate the infection speed, we define the relaxation time \(\tau\) as the time required for half of the particles to become infected. This threshold balances the need to capture meaningful infection dynamics while avoiding early-stage fluctuations and saturation effects that occur at very high infection levels Using this relaxation time, we can estimate the infection speed throughout the system, thus capturing the temporal dynamics of how the infection spreads. We examine how the relaxation time \(\tau\) is determined by the diffusion coefficient \(D\) and the infection radius \(r\) for a fixed infection rate of \(\beta = 0.05\), where the density is also fixed as \(\rho = N/L = 0.1\), with \(N = 10^2\) and \(L = 10^3\). The infection radius defines the range within which an infected particle can infect others, significantly impacting the spread of infection.

\subsection{The Role of Infection Radius in Diffusion-Controlled Systems}

\begin{figure*}
    \includegraphics[width=1.\textwidth]{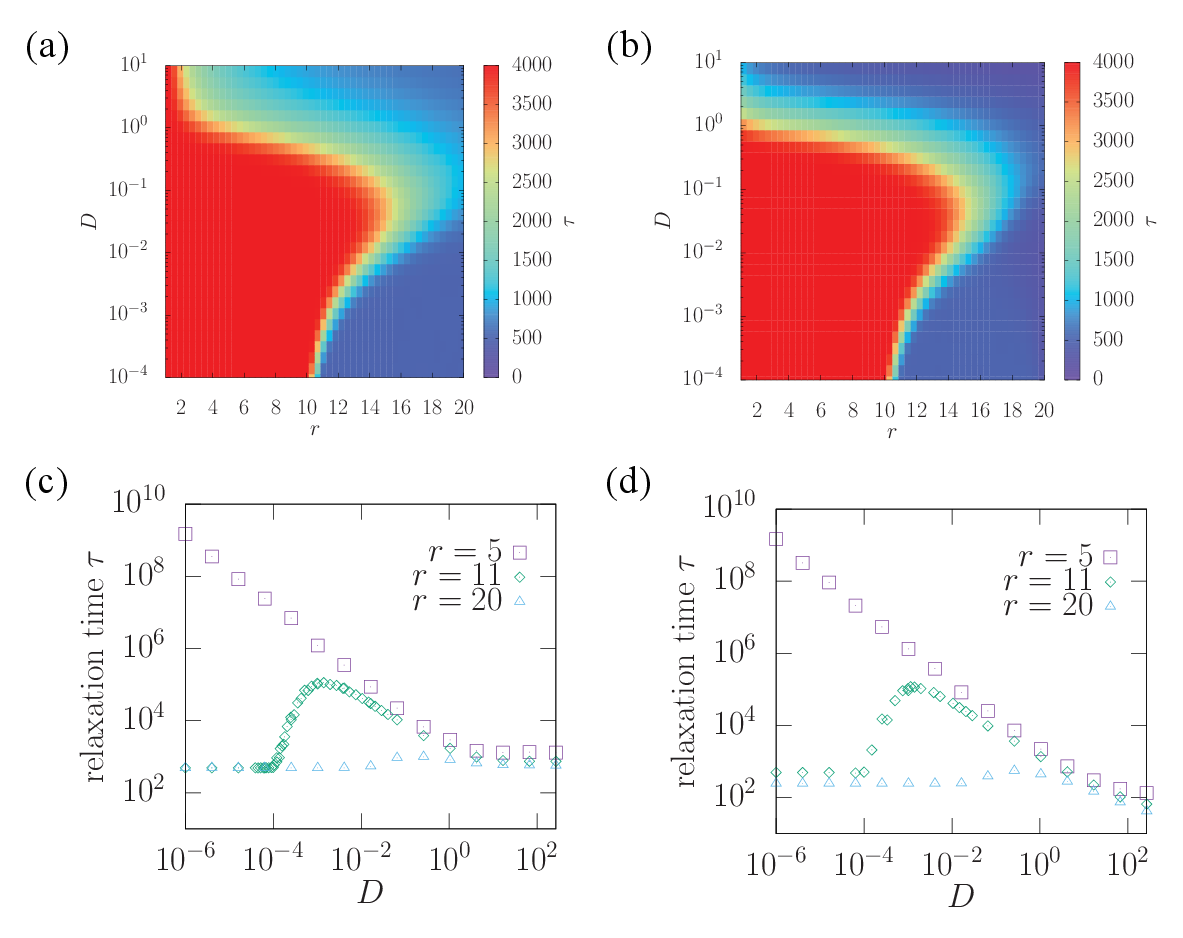} 
    \caption{
    Infection relaxation time \(\tau\)  with infection rate \(\beta = 0.05\), $L=10^3$, and $N=10^2$, where the initial particle configuration is a lattice arrange.  The number of the ensemble is $10^3$.
        (a) Heatmap of infection relaxation time  in hardcore repulsive system, 
        (b) Heatmap of infection relaxation time  in free-diffusive system, 
        (c) Infection relaxation time \(\tau\) as a function of the diffusion coefficient \(D\) in hardcore repulsive system for different values of $r$, and (d) Infection relaxation time \(\tau\) as a function of the diffusion coefficient \(D\) in free-diffusive system for different values of $r$. 
        The symbols are the results of numerical simulations. In the heatmaps, a cutoff is applied at  $\tau = 4000$  due to computational constraints. If the relaxation time exceeds 4000, its value is set to 4000 for numerical feasibility.
        }
    \label{fig:combined}
\end{figure*}

Our simulations of both hardcore repulsive and free-diffusive Brownian particle systems reveal a clear and consistent relationship between the infection radius \(r\) and the infection relaxation time \(\tau\). Figures~\ref{fig:combined}(a) and (b) show the heatmap of the relaxation time in \((r, D)\) space, demonstrating its dependence on both \(r\) and the diffusion coefficient \(D\). 
A cutoff at  $\tau = 4000$  is applied in the heatmaps due to computational constraints, meaning that any values exceeding this threshold are set to 4000 for numerical feasibility. In contrast, the values shown in Figs~\ref{fig:combined}(c) and (d) represent the actual infection relaxation times for different infection radii without truncation. As demonstrated in Figures~\ref{fig:combined}(c) and (d), the infection relaxation time increases exponentially as the diffusivity decreases, particularly when the infection radius is smaller than the mean spacing in the initial configuration. This trend highlights the strong influence of particle mobility on the spreading dynamics in {\color{black}low-infection-radius} regimes, where infection propagation is more sensitive to diffusive transport. 

For both hardcore repulsive and free-diffusive systems, the relaxation time \(\tau\) decreases monotonically with increasing \(r\) across all values of \(D\), with this trend being particularly pronounced in the low-diffusivity regime. 
{\color{black}Importantly, there is no qualitative difference in the  dependence of $\tau$  on  $r$   between the two cases; both exhibit a monotonically decreasing trend, regardless of the presence or absence of particle-particle interactions. This consistency arises because a larger infection radius allows each infected particle to reach more susceptible neighbors within its vicinity, thereby increasing transmission opportunities and accelerating the spread of infection in both systems.}

\subsection{Optimal Diffusivity for Infection Control}

Our simulations reveal the existence of an optimal diffusion coefficient that maximizes the infection relaxation time, effectively minimizing the infection spreading speed in both hardcore repulsive and free-diffusive systems. This counterintuitive behavior emerges from a delicate balance between particle mixing and the frequency of infection-inducing encounters.

{\color{black}To understand the conditions under which this optimal diffusivity emerges, we analyze the dependence of the relaxation time  $\tau$  on the diffusion coefficient  $D$  for different values of the infection radius  $r $. The results reveal two distinct regimes: in the first regime,  $\tau$  decreases monotonically with increasing  $D $, while in the second, a non-monotonic behavior appears, giving rise to a well-defined minimum in the infection speed.
}

In the first regime, where the infection radius $r$ is smaller than a threshold $r^*$, the relaxation time $\tau$ decreases monotonically with increasing $D$. Numerical simulations reveal that the threshold \(r^*\) is determined by the interparticle spacing of the initial lattice configuration, \(L/N\). This threshold radius is nearly identical across both hardcore repulsive and free-diffusive systems, approximately equal to  $r^*\approx L/N=10$ under the simulation conditions. This monotonically decreasing trend implies that higher diffusivity enhances the rate of infection spread by increasing the likelihood of contact between infected and susceptible particles. In this low-radius regime, limited contact opportunities dominate the dynamics, making diffusion essential for accelerating infection [see Figs.~\ref{fig:combined}(c) and (d)].

In contrast, when the infection radius $r$ exceeds the threshold $r^*$, a more complex behavior emerges. At zero diffusion ($D = 0$), the relaxation time is relatively short as infection progresses through static proximity (i.e., without particle movement). 
 However, in the low-diffusive regime, $\tau$ increases, suggesting that limited mobility temporarily hinders the spread of infection. As the diffusion coefficient increases further, the relaxation time decreases again and eventually converges. This behavior indicates the existence of an intermediate, optimal diffusion coefficient that maximizes the relaxation time. The non-trivial peak in $\tau$ reflects the intricate interplay between particle mixing and infection interactions [see Figs.~\ref{fig:combined}(c) and (d)]. 
Moreover, increasing the infection radius \(r\) shifts the peak of the relaxation time to a high diffusivity, indicating that the diffusion coefficient \(D\) at which the relaxation time is the longest becomes larger. This phenomenon further highlights the influence of the infection radius on particle dynamics and infection spreading. %These results underscore the importance of spatial parameters, particularly the infection radius, in controlling infection dynamics.

These findings highlight the complexity of infection dynamics and demonstrate that controlling diffusivity can significantly impact the spread of infections, with potential applications in designing effective containment strategies.

\subsection{Impact of Hardcore Interaction on Infection Dynamics}

Hardcore interactions suppress particle diffusivity and  influence the infection relaxation time in high-diffusivity regime. 
As shown in Figs.~\ref{fig:combined}(c) and (d), the infection relaxation time $\tau$ in free-diffusive systems exhibits behavior similar to that observed in hardcore repulsive systems. 
The threshold $r^*$, which marks the transition in infection dynamics, remains consistent between the two cases. 

For low-diffusivity regime, no significant difference in the relaxation time \(\tau\) is observed between hardcore repulsive and free-diffusive systems. This is because, in the low-diffusivity regime, when adjacent susceptible particles come into contact with infected particles, infection occurs within the infection radius before the susceptible particles pass the infected particles, resulting in the minimal influence of inter-particle interactions.
    
    At higher diffusion coefficients, the relaxation time $\tau$ asymptotically converges to a lower value in free-diffusive systems compared to hardcore repulsive systems. 
    Theoretical results for the high-diffusivity limit ($D = \infty$) confirm this trend (see Appendix~B). 
    This reflects the enhanced ability of fast diffusion to facilitate rapid mixing and infection spread when particles are not constrained by interactions [see Figs.~\ref{fig:combined}(c) and (d)]. Additionally, a leapover effect emerges in the infection dynamics, where susceptible particles bypass intermediate infected particles, further enhancing the infection speed in free-diffusive systems.

These findings underscore the significant role of hardcore interactions in moderating infection dynamics. Physical constraints in hardcore repulsive systems suppress the spread of infection by limiting it to sequential interactions. In contrast, free-diffusive systems achieve faster mixing and propagation under high-diffusivity conditions, promoting rapid infection spread by enhancing particle mobility and mixing.

\section{Discussion}

\subsection{Understanding the Emergence of Optimal Diffusivity Through Diffusive Timescales}

\begin{figure}
    \includegraphics[width=0.45\textwidth]{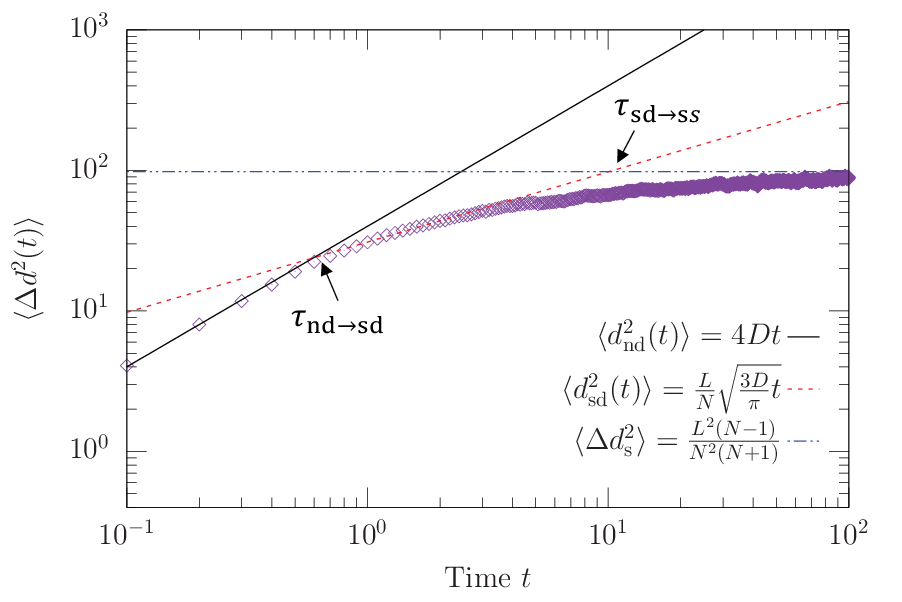}
    \caption{Mean square change in the distance  of neighboring particles as a function of time in a single-file system, where the hardcore repulsive interactions are present. Symbols are the results of numerical simulations. The simulation parameters are \(N=100\), \(L=1000\), \(D=10\) with integration time step of \(dt=0.01\), where  the initial configuration of the particles is a lattice arrangement and  the number of the ensemble is $10^4$. Three lines  correspond to theoretical predictions: $\langle d^2_\mathrm{nd} (t)\rangle = 4Dt$ (normal diffusion), $\langle d^2_\mathrm{sd} (t)\rangle = \frac{L}{N}\sqrt{\frac{3D}{\pi}t}$ (subdiffusive behavior in single-file diffusion), and $\langle \Delta d^2_{\mathrm{s}}\rangle=\frac{L^2(N-1)}{N^2(N+1)}$ (theoretical value in the steady state). }
    \label{fig: msd}
\end{figure}

\begin{figure}
    %\centering
    \includegraphics[width=.45\textwidth]{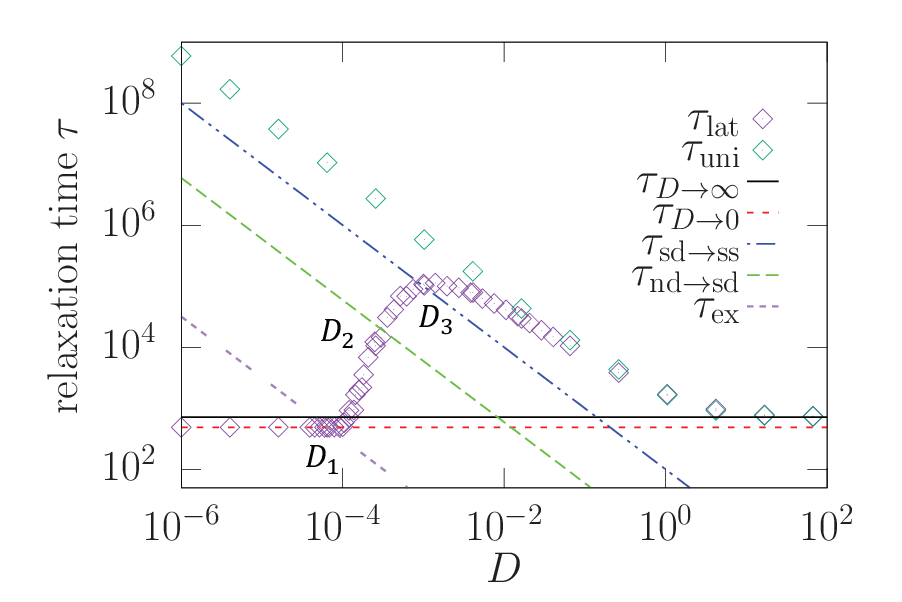} 
    \caption{
    Initial configuration dependence of infection relaxation time in the hardcore repulsive system, where $r=11$ and all other parameters are the same as in Fig.~\ref{fig:combined}, The ensemble size is $10^3$.
    The relaxation times $\tau_{\text{lat}}$ and $\tau_{\text{uni}}$ correspond to the infection relaxation time calculated for the initial configurations with lattice configuration and  the uniform distribution, respectively. 
    Solid lines represent the theoretical predictions of the relaxation times $\tau_{D \to 0}$ and $\tau_{D \to \infty}$ for low and high diffusive limit, respectively. Three lines represent three characteristic time scales $\tau_{\mathrm{ex}}, \tau_{\mathrm{nd \to sd}}$ and $\tau_{\mathrm{sd \to ss}}$ as a function of $D$, corresponding to the transitions described by Eqs.~(\ref{eq: rel time ex}), (\ref{eq main:nd_to_sd}) and (\ref{eq main:sd_to_ss}), respectively.
    }
    \label{fig:infection_relaxation_initial}
\end{figure}

We investigate how optimal diffusivity emerges for $N/L<r$ in both hardcore repulsive and free-diffusive systems by analyzing three characteristic timescales of diffusion dynamics and comparing them with the infection relaxation time. To characterize the diffusive characteristic timescales, we examine the mean square change in the distance  (MSCD) of neighboring particles,  defined as $\langle \Delta d^2 (t) \rangle$, where $\Delta d^2 (t)= (d(t) -d(0))^2$ and $d(t)=x_{i+1}(t) -x_{i}(t)$. As shown in Fig.~\ref{fig: msd}, the MSCD in a single-file system transitions over time from normal diffusion to subdiffusion and eventually to a steady state. 

In hardcore repulsive systems, we define two characteristic timescales to describe the transitions in particle dynamics. The transition time from normal diffusion to subdiffusion,  $\tau_{\mathrm{nd \to sd}} $, which marks the point at which particles begin to experience hardcore repulsive interactions, resulting in subdiffusive behavior. The crossover time from subdiffusion to the steady state,  $\tau_{\mathrm{sd \to ss}}$, which represents the characteristic relaxation time required for the particle configuration to reach equilibrium. In free-diffusive systems, these characteristic timescales still exist but have different physical interpretations. The timescale $\tau_{\mathrm{nd \to sd}} $  corresponds to the point at which particles begin to experience overlaps, rather than hardcore interactions. Meanwhile, $\tau_{\mathrm{sd \to ss}}$ represents the relaxation time for the system to reach equilibrium, which, when ignoring particle numbering, is equivalent to the hardcore repulsive case.

These timescales are determined by identifying the crossover points in the MSCD.
 These crossover points correspond to the intersections of theoretical MSCD curves for normal diffusion ($\langle d^2_\mathrm{nd}(t) \rangle = 4Dt$), subdiffusion ($\langle d^2_\mathrm{sd}(t) \rangle = \frac{L}{N} \sqrt{\frac{3D}{\pi}} \sqrt{t}$), and the steady state  (see Appendix~C). The characteristic timescales are given by the following expressions:
\begin{equation}
\tau_{\mathrm{nd \to sd}} = \frac{3}{16\pi   \rho^2 D},
\label{eq main:nd_to_sd}
\end{equation}
and
\begin{equation}
\tau_{\mathrm{sd \to ss}} = \left\{\frac{N-1}{ \rho(N+1)}\right\}^2 \frac{\pi}{3D} .
\label{eq main:sd_to_ss}
\end{equation}
In the large  $N$  limit, where the density  $\rho = N/L$  remains fixed, these characteristic timescales approach constant values and become independent of  $N$. Consequently, they do not depend on the system size.
Both timescales are inversely proportional to the diffusion coefficient ($D^{-1}$), highlighting a clear physical relationship between diffusion and relaxation dynamics. This dependence emphasizes how increased diffusivity accelerates the transition between different transport regimes, affecting the overall infection spreading behavior.

Since infection spreads through pairwise interactions between susceptible and infected particles, the infection relaxation time is influenced by whether a particle escapes the infection radius before transmission occurs. If one of the  $N/2$  particles surrounding the initially infected particle at the center moves beyond the infection radius before becoming infected, the infection process is temporarily delayed, impacting the overall infection relaxation time.
 To quantify this effect, we introduce a characteristic escape time, related to the mean fastest first passage time (MFFPT). This time describes the time required for distance between two Brownian particles, among $N/2$ independent Brownian particles initially separated by $x_{i+1}(0) -x_{i}(0)= L/N$, to reach the threshold $|x_{i+1}(t) -x_{i}(t)| = r$. The MFFPT is derived in \cite{weiss1983order,yuste2000diffusion,yuste2001order,yuste2001ordertrap} as
\begin{equation}
\tau_{\mathrm{ex}} = \frac{(r -  \rho^{-1})^2}{8D \ln(N/2)}.
\label{eq: rel time ex}
\end{equation}
The MFFPT is a decreasing function of $N$, meaning that as the number of particles increases (while keeping density fixed), the minimum escape time becomes shorter. This suggests that the system size influences the infection process, even though the density remains constant. The faster escape of particles in larger systems increases the probability that at least one particle escapes the infection radius before transmission occurs, causing the onset of the increase in the infection relaxation time to shift to a lower diffusivity. 
Although there are two targets located at $x_{i+1}(t) -x_{i}(t)=r$ and $x_{i+1}(t) -x_{i}(t)=-r$, the fastest particle's passage time is primarily dictated by the closest boundary at $x_{i+1}(t) -x_{i}(t) = r$, as the influence of the more distant target is negligible.
This escape time is also inversely proportional to the diffusion coefficient ($D^{-1}$), emphasizing the role of diffusivity in determining how quickly two particles can separate to a critical distance. 

The infection relaxation time for $L/N < r$ exhibits a strong dependence on the initial particle configuration, especially in the low-diffusivity regime. Figure~\ref{fig:infection_relaxation_initial} shows how the infection relaxation time varies with the diffusion coefficient for different initial conditions in the hardcore repulsive systems, where  $\tau_{\mathrm{lat}}$  represents the infection relaxation time for an initial  configuration in which particles are arranged in a uniform lattice with equal spacing  $L/N$, and $ \tau_{\mathrm{uni}}$  corresponds to the infection relaxation time for a random initial configuration, in which particles' positions are uniformly distributed. When the initial configuration is a steady state (uniform distribution), the infection relaxation time decreases significantly as $D$ increases. In this scenario, higher diffusivity enhances particle-particle encounters, thereby accelerating the spread of infection.
In contrast, when the initial configuration is a lattice arrangement, the infection relaxation time shows a more complex dependence on $D$. Specifically, in the low-diffusivity regime, the infection relaxation time is much smaller than the particle diffusion relaxation time $\tau_{\mathrm{sd \to ss}}$, indicating that infection dynamics are dominated by the initial configuration. As shown in Fig.~\ref{fig:infection_relaxation_initial}, the behavior of the infection relaxation time  $\tau_{\mathrm{lat}}$  for  a lattice initial configuration with $L/N < r$  exhibits transitions characterized by three transition diffusivities—$D_1, D_2$, and $D_3$.

\begin{enumerate}
	\item \textbf{First Transition Point $D_1$ (Onset of Slower Infection Spreading)}: In the low-diffusivity regime, the escape time, i.e., the time for a particle to move beyond the infection radius, is much shorter than the infection relaxation time. As a result, the infection relaxation time initially remains unchanged even as the diffusivity increases. However, as diffusivity increases further, some infected particles escape before transmission occurs, leading to a prolongation of the infection relaxation time. At $D_1$, the infection relaxation time $\tau_{\mathrm{lat}}$ begins to increase with $D$. Below $D_1$, the infection relaxation time remains nearly constant, showing little dependence on $D$. This point corresponds to the condition where the infection relaxation time matches the MFFPT $\tau_{\mathrm{ex}}$ for $N/2$ particles starting at $L/N$. When the infection relaxation time  exceeds $\tau_{\mathrm{ex}}$, a ``link" in the infection chain breaks, causing a rise in the infection relaxation time. The diffusion coefficient $D_1$ can be derived as the intersection between $\tau_{\mathrm{ex}}$ and $\tau_{D \to 0}$ (see Appendix~A):
\begin{equation}
D_1 = \frac{(r - \rho^{-1})^2 \beta}{2 \ln(N/2)(N - 2)}.
\end{equation}
This equation shows that $D_1$ depends quadratically on the infection radius $r$; as $r$ increases, $D_1$ also increases quadratically due to the longer escape time for susceptible particles. 
Moreover, $D_1$ decrease as the system size $ L$  or the number of particles  $N$  increases, while keeping the density  $\rho$  fixed. Consequently, in the thermodynamic limit, this transition disappears,  and the infection relaxation time increases immediately as  $D$  increases. As shown in Fig.~\ref{fig:infection_relaxation_initial}, the theoretical prediction aligns well with the simulation results. Notably, this regime is not influenced by particle interactions, indicating that the observed transition is purely a result of diffusive transport effects rather than direct interactions between particles.

	\item	\textbf{Second Transition Point $D_2$ (Onset of Moderate Increase of the Infection Relaxation Time)}: Beyond $D_1$, as diffusivity increases, more particles escape the initial infection region, temporarily slowing the infection rate and increasing the infection relaxation time. As diffusivity increases further, the infection relaxation time becomes larger than $\tau_{\mathrm{nd \to sd}}$, the time scale associated with the transition from normal diffusion to subdiffusion. In this regime, the confinement imposed by particle interactions reduces the escape rate, leading to a moderate increase of the infection spreading speed. This moderation begins  at $D_2$. For $D>D_2$, particle interactions become significant, further slowing the growth of the infection relaxation time. This phase represents a moderate increase in the infection relaxation time, distinguishing it from the previous rapid-growth regime observed for  $D < D_2 $. As shown in Fig.~\ref{fig:infection_relaxation_initial}, this moderation begins when the infection relaxation time matches $\tau_{\mathrm{nd \to sd}}$.  As diffusivity increases further, the infection relaxation time reaches its maximum at the transition diffusivity $D_3$, marking the boundary between this increasing phase and the subsequent decreasing regime. While we discuss the role of particle interactions in this phase, it is important to note that this moderate phase persists even in the absence of interactions, as overtaking effects introduce a form of confinement that similarly influences infection spreading dynamics.
	
	\item	\textbf{Third transition point $D_3$ (Peak of the Infection Relaxation Time)}: At high diffusivity, particles mix rapidly, allowing susceptible individuals to frequently encounter infected ones. This restores fast transmission dynamics, leading to a decrease in the infection relaxation time as infection spreads more efficiently. Beyond $D_3$, as diffusivity increases, the infection relaxation time begins to decrease with  $D$. In this regime, the particle configuration relaxation time $\tau_{\mathrm{sd \to ss}}$ becomes comparable to the infection relaxation time. As a result, for $D > D_3$, the infection relaxation time is no longer influenced by the initial configuration, leading to a decrease in relaxation time and the emergence of a peak. 
The emergence of the peak at  $D_3$  is driven by two competing effects: On one hand, particle escape delays infection spreading, leading to an increase in the infection relaxation time. On the other hand, as diffusivity increases further, faster mixing compensates for escape effects, resulting in more frequent encounters and a shorter infection relaxation time. 
Thus,  $D_3$  marks the transition point where the infection relaxation time shifts from being dominated by escape dynamics to being controlled by rapid mixing and frequent encounters.
As shown in Fig.~\ref{fig:infection_relaxation_initial}, the theoretical prediction aligns well with the simulation results. Notably, since the MSCD is identical in both hardcore repulsive and free-diffusive systems, the particle configuration relaxation time $\tau_{\mathrm{sd \to ss}}$ remains unchanged. While the mechanisms of infection spreading differ between hardcore repulsive and free-diffusive systems, the third transition point is consistent across both systems.
\end{enumerate}

These results highlight the intricate interplay between diffusion dynamics, particle interactions, and initial configurations in determining infection relaxation times. By identifying these key transition points, we provide a robust framework for understanding optimal diffusivity in controlling infection speed.

\subsection{Thermodynamic Limit and Finite-Size Effects}
Finally, we discuss finite-size effects. As shown in Eqs. \eqref{eq main:nd_to_sd} and \eqref{eq main:sd_to_ss}, the characteristic diffusion timescales $\tau_{\mathrm{nd \to sd}}$ and 
$\tau_{\mathrm{sd \to ss}}$ do not depend on the system size $L$ or the number of particles $N$ in the thermodynamic limit $L\to \infty$ and $N\to \infty$ while keeping the density $N/L$ fixed.  
As shown in Appendix~A,   the infection relaxation time  $\tau_{\mathrm{lat}}$  for  a lattice initial configuration with $L/N < r$ and $D=0$ is a monotonically increasing function of $N$. 
Thus, as $N$ increases,  $\tau_{\mathrm{lat}}$ at $D=0$ also increases. Since $\tau_{\mathrm{sd \to ss}}$  does not depend on $N$ in the thermodynamic limit, there exists a diffusion coefficient $D_*$ such that the infection relaxation time  $\tau_{\mathrm{lat}}$ at $D=D_*$ matches $\tau_{\mathrm{sd \to ss}}$ at the same $D_*$. Because the infection relaxation time decreases with increasing $D$ once the configuration has equilibrated, this diffusivity $D_*$ corresponds to the transition point above which the infection relaxation time decreases. Consequently, the optimal diffusivity that maximizes the infection relaxation time always exists when $L/N < r$.  Theoretically, this behavior persists for arbitrarily large systems as long as the system size remains finite. However, in the thermodynamic limit, where both the number of particles and the system size tend to infinity while keeping the density fixed, the optimal diffusivity disappears because $D_*$ becomes zero. This indicates that the emergence of optimal diffusivity is inherently a finite-size effect.
We conducted numerical simulations for different system sizes and confirmed the universality of the optimal diffusivity for infection control.

\subsection{Beyond 1D: Infection Spreading in Higher Dimensions}
The slowdown of infection spreading due to hardcore repulsive interactions is a characteristic feature of 1D systems, where particles cannot overtake each other, leading to single-file diffusion. In higher dimensions, particles can bypass one another, reducing the impact of hardcore repulsion on infection dynamics. Despite this difference, we confirmed that the emergence of an optimal diffusivity persists in 2D systems as well, suggesting that the observed non-monotonic dependence of the infection relaxation time on diffusivity is a general phenomenon in diffusion-driven epidemic spreading.
Additionally, 2D systems can be effectively mapped onto 1D free-diffusive systems by modifying the infection rate, accounting for infection opportunities in the transverse direction. This conceptual connection supports the use of 1D models as a foundation for understanding more complex spreading dynamics in quasi-1D confined environments, such as narrow corridors, microfluidic channels, and biological transport systems.

\section{Conclusion}

We performed numerical simulations to unveil 
how the infection radius and particle diffusivity significantly affect the infection spreading. When the infection radius is smaller than $L/N$, the infection relaxation time decreases monotonically with increasing diffusion coefficient, as higher diffusivity enhances particle encounters and accelerates transmission.
In contrast, when the infection radius exceeds $L/N$, the infection relaxation time exhibits a nontrivial peak, indicating the presence of an optimal diffusivity. 
{\color{black}This behavior emerges under a non-equilibrium initial configuration in which particles are uniformly spaced at regular intervals exceeding the infection radius. In such structured setups,}
the emergence of this optimal diffusivity  arises from a trade-off between contact frequency and escape probability: while higher diffusivity increases encounters, it also allows susceptible particles to avoid infection more effectively. As a result, a specific diffusivity most effectively suppresses transmission.
{\color{black}Importantly, our results show that hardcore interactions do not qualitatively affect the relaxation time in either regime, and the optimal diffusivity appears in both interacting and non-interacting systems.} 
These findings offer a conceptual framework for designing infection control strategies based on mobility regulation. By tuning movement and contact rates, it may be possible to exploit this trade-off and minimize the overall spread.
Our simulations also suggest that this optimal diffusivity remains robust across system sizes, highlighting the generality of the mechanism{\color{black}---provided the initial spatial structure is present.}

\appendix
\section*{Appendix}

\section{Infection dynamics in the low-diffusive limit (\(D = 0\)) in hardcore repulsive systems}
\label{sec:zero_diffusion}

\begin{figure}
    \centering
    \includegraphics[width=0.45\textwidth]{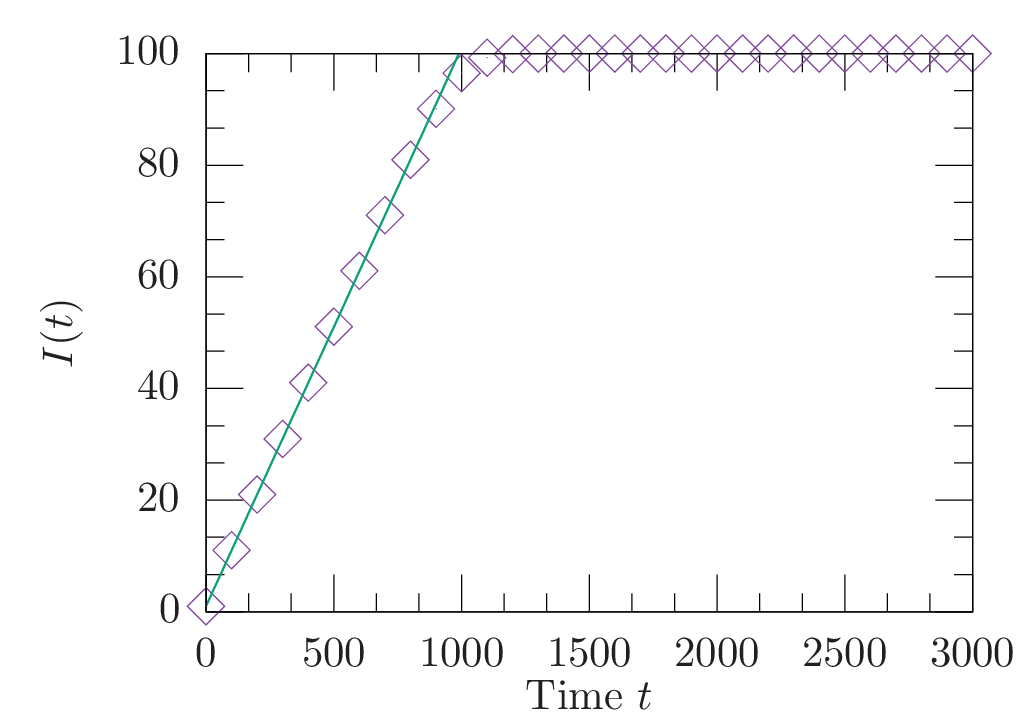}
    \caption{
    Time evolution of the number of infected particles \(I(t)\) in the static limit (\(D=0\)), with \(N=100\), \(L=1000\), $\beta= 0.05$ and the initial state arranged on a lattice. The graph compares simulation results with the theoretical prediction given by Eq.~\eqref{eq:I(t)D=0}. The number of the ensemble in numerical simulations is $10^5$.
    }
    \label{fig:td_D0}
\end{figure}

In the low-diffusive limit, infected particles remain fixed and do not move. Once an infected individual appears in the system, there is always a constant number of susceptible particles within the infection radius throughout the entire system. Specifically, since susceptible particles remain within a range of two particle lengths from the infected, the number of susceptible particles within the infection radius is always 2.

The time evolution of the probability \( p_k(t) \) that there are \( k \) infected particles at time \( t \) is given by the following master equation \cite{stollenwerk2000master,sanchez2021modelling}:
\begin{equation}
    \begin{cases}
    \dfrac{dP_1(t)}{dt} & = -2{\beta}P_1(t) \\
    \dfrac{dP_k(t)}{dt} & = 2{\beta}P_{k-1}(t)-2{\beta}P_{k}(t) \text{ ($2\leq k\leq N-1$)}\\
    \dfrac{dP_N(t)}{dt} & = 2{\beta}P_{N-1}(t) .
    \end{cases}
\end{equation}
The equations represent the dynamic process of the number of infected particles increasing or decreasing, allowing us to analyze the changes in the number of infected over time. 
 Solving the master equation yields 
\begin{equation}
    \begin{cases}
    P_k(t) & = \dfrac{(2\beta t)^{k-1}}{(k-1)!} e^{-2\beta t} \text{ \quad ($1\leq k\leq N-1$)}\\
    P_N(t) & = 1 - \sum_{k=1}^{N-1} P_{k}(t) .
    \end{cases}
\end{equation}
The mean number of infected people at time $t$ is given by
\begin{equation}
    \langle I(t) \rangle = \sum_{k=1}^{N} k P_{k}(t).
\end{equation}
For $N\gg 1$, $\langle I(t) \rangle$ can be approximated as
\begin{equation}
    \langle I(t) \rangle \approx 2\beta t + 1.
    \label{eq:I(t)D=0}
\end{equation}
This approximated result is in a good agreement with numerical simulations (see Fig.~\ref{fig:td_D0}).
From this result, the infection relaxation time \(\tau_{D \to 0}\) is obtained as 
\begin{equation}
    \tau_{D \to 0} = \frac{N-2}{4\beta}.
\end{equation}
This result is exact, despite the use of an approximation in the expression for $\langle I(t) \rangle$, as the infection dynamics follow a Poisson process.

\section{Infection dynamics in the high-diffusive limit (\ensuremath{D = \infty}) in the hardcore repulsive system}

{\color{black}
In the high-diffusivity limit, particle positions are sampled from a uniform distribution over the interval [0, L]. In the presence of hardcore interactions, the particle order remains fixed over time due to the no-overtaking constraint. In our numerical simulations, we resample particle positions from the uniform distribution at each time step. When hardcore interactions are present, particles are re-ordered from left to right after each resampling to preserve their original indexing consistent with single-file dynamics. In contrast, in the absence of hardcore interactions, the particle indices are maintained regardless of spatial position. In the hardcore case, this procedure yields a steady-state configuration characterized by a uniform spatial distribution with fixed particle order.}
The PDF \(P(d)\) of the distance \(d\) between two neighboring particles in a single-file diffusion system can be expressed as 
\begin{equation}
P(d) = \frac{N-1}{L} \left(1 - \frac{d}{L}\right)^{N-2},
\label{eq:pdf}
\end{equation}
where \(N\) is the number of particles, and \(L\) is the total system length. The shape of this distribution becomes steeper as the number of particles \(N\) increases, reflecting the higher particle density.
Using this PDF, the probability that the distance \(x\) between any two adjacent particles  is within the infection radius \(r\) can be computed. This probability determines the average number of susceptible particles \(\langle n_\mathrm{s} \rangle\) in contact with an infected particle, which is given by
\begin{equation}
        \langle n_\mathrm{s} \rangle = 2\int_{0}^{r}P(x)dx %\quad (1 \leq i \leq N-1)\\
        = 2\left\{1-\left(1-\frac{r}{L}\right)^{N-1}\right\}.
\end{equation}
For $N \gg 1$, $\langle n_s \rangle$ asymptotically approaches 2, indicating that, on average, there are two susceptible particles within the infection radius at the boundary of the infected cluster.

\begin{figure}
    \centering
    \includegraphics[width=0.45\textwidth]{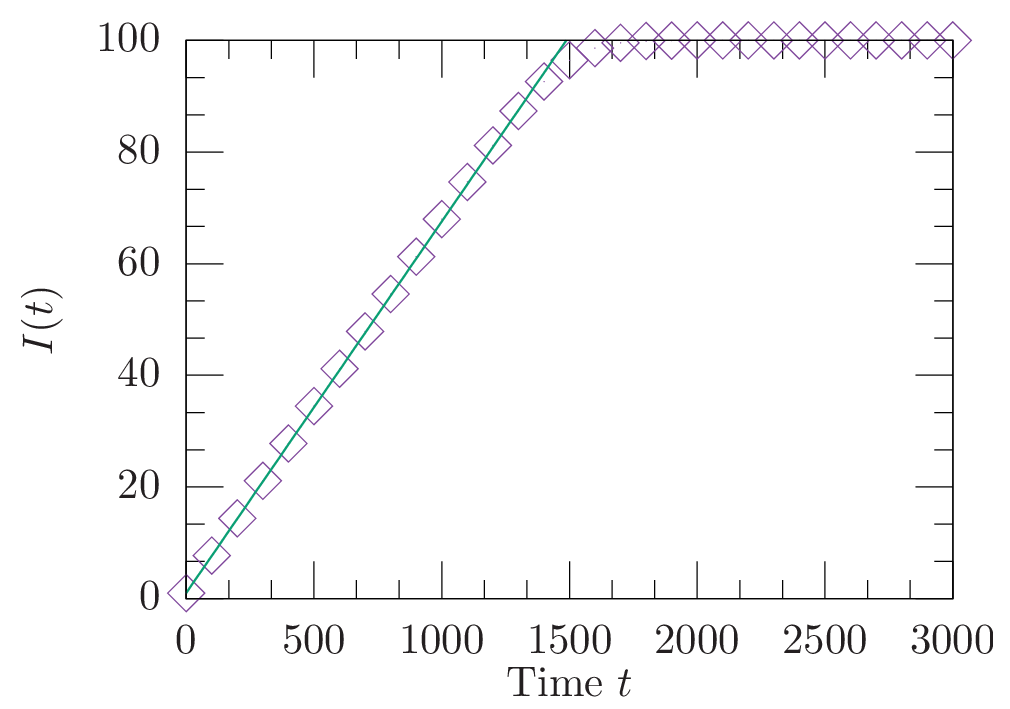}
    \caption{
    Time evolution of the number of infected particles \(I(t)\) in the hardcore repulsive system (\(D=\infty\)), with \(N=100\), \(L=1000\), and the initial state arranged on a lattice. The graph compares simulation results with the theoretical prediction given by Eq.~\eqref{eq:I(t)D=inf_con}. In the numerical simulations, particle positions are sampled from a uniform distribution at all times, while their individual ordering remains unchanged. The number of the ensemble in numerical simulations is $10^5$.
    }
    \label{fig:td_con}
\end{figure}

With this result, the master equation for the number of infected particles $I(t)$ in the high-diffusive limit can be expressed as
\begin{equation}
\frac{\mathrm{d}P_k(t)}{\mathrm{d}t} = \beta \langle n_s \rangle \left[P_{k-1}(t) - P_k(t)\right],
\end{equation}
where $\beta$ is the infection rate and $P_k(t)$ is the probability of having $k$ infected particles at time $t$. Solving this equation using the same approach as in Appendix~A yields the time evolution of the expected number of infected particles as
\begin{equation}
    \langle I(t) \rangle = 2\beta \left\{1 - \left(1 - \frac{r}{L}\right)^{N-1} \right\} t + 1.
    \label{eq:I(t)D=inf_con}
\end{equation}
This result is in a good agreement with numerical simulations (see Fig.~\ref{fig:td_con}). From this result, the infection relaxation time \(\tau_{D \to \infty}\) is obtained as
\begin{equation}
    \tau_{D \to \infty} = \frac{N - 2}{4\beta \left\{1 - \left(1 - \frac{r}{L}\right)^{N-1} \right\}}.
\end{equation}
This expression provides an analytical representation of the temporal evolution of the number of infected particles in the high-diffusive limit, taking interactions into account. Including interactions between particles is predicted to significantly change the speed and pattern of infection progression.

\section{Theory of interparticle distance in single-file systems}
\label{sec:single_file}
% Add content here related to the theory of interparticle distance in single-file diffusion systems.

\subsection{Mean square change in distance of  two neighboring particles in the steady state}

%In the steady state, the PDF \(P(d)\) of the distance \(d\) between two neighboring particles in a single-file diffusion system can be expressed as $P(d)=f_{i,i+1}(d)$:
%\begin{equation}
%\color{black}
%P(d) = \frac{N-1}{L} \left(1 - \frac{d}{L}\right)^{N-2},
%\label{eq:pdf}
%\end{equation}
%where \(N\) is the number of particles, and \(L\) is the total system length. This equation assumes that the particles are uniformly distributed. The shape of this distribution becomes steeper as the number of particles \(N\) increases, reflecting the higher particle density.

In the long-time limit, the mean square change in the distance (MSCD) between two neighboring particles converges to the variance of their distance in the steady state. 
Using the PDF of interparticle distance \(P(d)\), i.e., Eq.~(\ref{eq:pdf}), the variance of the distance can be calculated as
\begin{equation}
\langle \Delta d_\mathrm{s}^2 \rangle = \int_{0}^{L} \left(x - \frac{L}{N}\right)^2 P(x) \, dx,
\label{eq:msd_integral}
\end{equation}
where \(d - \frac{L}{N}\) represents the deviation of the particle distance \(d\) from the mean particle spacing \(\frac{L}{N}\). %By integrating the squared deviation weighted by the PDF \(P(d)\), the MSDNP is obtained. 
Performing this integration yields
\begin{equation}
\langle \Delta d_\mathrm{s}^2 \rangle = \frac{L^2 ( N-1 )}{N^2 (N+1)}.
\label{eq:msd_result}
\end{equation} 
This result shows that the MSCD depends on both the number of particles \(N\) and the system length \(L\). As \(N\) increases, the fluctuations in the distance between neighboring particles decrease, indicating that the system becomes more tightly packed. This theoretical result serves as a reference value for the steady-state behavior, which can be compared with numerical simulation results.

\subsection{Dynamics of two-particle distance in single-file systems and transition points}

The dynamics of the distance between two particles in single-file diffusion systems exhibit distinct behaviors depending on the time scale. This subsection examines the theoretical transition points from normal diffusion to subdiffusion and from subdiffusion to the steady state.
%\subsubsection{Transition from normal diffusion to subdiffusion}
%In single-file diffusion, particles exhibit normal diffusion at short time scales, as they are not yet influenced by the spatial constraints of the single-file structure. The mean square displacement (MSD) of a single particle in normal diffusion is given by \cite{einstein1905motion}
%\begin{equation}
%\langle x^2 \rangle = 2Dt,
%\label{eq:normal_diffusion_msd}
%\end{equation}
%where \(D\) is the diffusion coefficient. 

%\section{MSD of the Distance Between Two Neighboring Particles}
The mean square distance between two neighboring particles can be expressed as
\begin{equation}
    \langle d^2 (t)\rangle = \langle (x_{i+1}(t) - x_i(t)  )^2 \rangle = 2\langle x_i^2 (t) \rangle - 2\langle x_{i+1}(t) x_i (t)\rangle,
\end{equation}
where  \(d (t) = x_{i+1}(t) - x_i(t) \)  is the distance between two neighboring particles and \(\langle x_{i+1}(t) x_i(t) \rangle\) represents the correlation function between neighboring particles. At the initial stages of diffusion, this correlation is negligible. Therefore, in the short-time regime, the motion remains normal diffusion, as the particles have not yet interacted strongly. Therefore, the MSCD in the initial stage of diffusion can be approximated as
\begin{equation}
 \langle  \Delta d^2 (t)\rangle \approx    \langle  d^2_\mathrm{nd} (t)\rangle = 4Dt,
    \label{eq:relative_normal_diffusion}
\end{equation}
where \(D\) is the diffusion coefficient.

As time progresses, particles begin to experience the effects of the single-file constraint, causing their motion to transition from normal diffusion to subdiffusion. This phenomenon is particularly noticeable in systems initialized with a lattice configuration. In this subdiffusive regime, the MSD of a single particle is known to asymptotically follow subdiffusion \cite{lizana2010foundation,sorkin2024uphill,leibovich2013everlasting}:
\begin{equation}
    \langle (x_{i}(t) - x_i(0))^2\rangle \sim  \frac{L}{N} \sqrt{\frac{2D}{\pi}} \sqrt{t},
    \label{eq:subdiffusion_msd}
\end{equation}
where \(L\) is the system length and \(N\) is the number of particles. This  subdiffusive behavior reflects the impact of particle interactions in the single-file system. During this stage, the correlation function $\langle x_{i+1}(t)x_i(t) \rangle$ between neighboring particles becomes significant and cannot be neglected in calculating the MSCD. Numerical simulations reveal that the MSCD grows proportionally to $\sqrt{t}$ in this regime. Fitting the MSD using the form of Eq.~(\ref{eq:subdiffusion_msd}) shows that the coefficient of the diffusion term is approximately 3. Therefore, the MSCD in the subdiffusive regime is expressed as
\begin{equation}
    \langle  \Delta d^2 (t)\rangle \approx \langle d^2_\mathrm{sd}(t) \rangle = \frac{L}{N} \sqrt{\frac{3D}{\pi}} \sqrt{t}.
    \label{eq:relative_subdiffusion}
\end{equation}
This subdiffusive behavior is transient, as the MSCD eventually converges to a constant value determined by the steady state of the system.

Using Eqs.~(\ref{eq:relative_normal_diffusion}) and (\ref{eq:relative_subdiffusion}), the crossover time from normal diffusion to subdiffusion is given by
\begin{equation}
\tau_{\mathrm{nd \to sd}} = \left(\frac{L}{N}\right)^2 \frac{3}{\pi}\frac{1}{16D}.
\label{eq:nd_to_sd}
\end{equation}
This equation indicates that the crossover time depends on the system size \(L\), the number of particles \(N\), and the diffusion coefficient \(D\). As \(N\) increases, the crossover occurs at shorter time scales, and a larger \(D\) further accelerates this transition. These results highlight the shift from independent motion to motion constrained by the single-file structure.

%\subsubsection{Transition from Subdiffusion to the Steady State}

As time progresses further, the system eventually reaches a steady state. The MSCD in the steady state is theoretically given by Eq.~(\ref{eq:msd_result}).
%\begin{equation}
%\langle d_\mathrm{s}^2 \rangle = \frac{L^2 \{ 2 + N(N-1) \}}{N^2 (N+1)(N+2)} .
%\label{eq:steady_state_msd}
%\end{equation}
From this, the diffusion relaxation time for the transition from subdiffusion to the steady state is obtained as
\begin{equation}
\tau_{\mathrm{sd \to ss}} = \left\{ \frac{L(N-1)}{N(N+1)}\right\}^2  \frac{\pi}{3D}.
\label{eq:sd_to_ss}
\end{equation}
This result indicates that the relaxation time depends strongly on the system size $L$, particle number $N$, and diffusion coefficient $D$. Specifically, as \(N\) increases, the time required to reach the steady state becomes shorter, while a larger \(D\) shortens the relaxation time.

%\bibliographystyle{unsrt}
%\bibliography{reference}

\end{document}